\documentclass[12pt]{article}
\usepackage{a4wide,amssymb,cite}
\usepackage{a4wide,amssymb,cite,graphicx}
\usepackage{epsfig}
\usepackage[usenames,dvipsnames]{color}
\usepackage{slashed}

\newcommand{\be}{\begin{equation}}
\newcommand{\ee}{\end{equation}}
\newcommand{\bea}{\begin{eqnarray}}
\newcommand{\eea}{\end{eqnarray}}

\def\circa#1{\,\raise.3ex\hbox{$#1$\kern-.75em\lower1ex\hbox{$\sim$}}\,}

\begin{document}

\begin{titlepage}
%
%


%

\begin{centering}
\vspace{1cm}
{\large {\bf Magnetic dark matter for the X-ray line at 3.55 keV}} \\

\vspace{1.5cm}

{\bf Hyun Min Lee}
\\
\vspace{.5cm}

{\it Department of Physics, Chung-Ang University, Seoul 156-756, Korea.} 
\\

\end{centering}
\vspace{2cm}

\begin{abstract}
\noindent
We consider a decaying magnetic dark matter explaining the X-ray line at $3.55\,{\rm keV}$ shown recently from XMM-Newton observations. We introduce two singlet Majorana fermions that have almost degenerate masses and fermion-portal couplings with a charged scalar of weak scale mass. In our model, an approximate $Z_2$ symmetry gives rise to a tiny transition magnetic moment between the Majorana fermions at one loop. The heavier Majorana fermion becomes a thermal dark matter due to the sizable fermion-portal coupling to the SM charged fermions.
We find the parameter space for the masses of dark matter and charged scalar and their couplings, being consistent with both the relic density and the X-ray line. Various phenomenological constraints on the model are also discussed.

\end{abstract}

\vspace{3cm}

\end{titlepage}

\section{Introduction}

Dark matter is a dominant component of the matter density in the universe, playing a crucial role in the structure formation and explaining the flatness of galaxy rotation curves, etc.  The evidences for dark matter are without question but there is no understanding of the properties of dark matter such as its mass or coupling to the SM particle, except gravitational interaction. 

Recently, there was an interesting indication of dark matter  \cite{xray} from the stacked X-ray spectrum of galaxies and clusters \cite{xmm}, which shows an unexplained line signal at the energy $3.55\,{\rm keV}$. A sterile neutrino having a very small mixing with the active neutrinos, can explain the X-ray line signal by a small transition magnetic moment but the link to the generation of neutrino masses via see-saw mechanism is unclear due to a small mixing \cite{xray}. The model building issue with keV sterile neutrino was discussed in Ref.~\cite{keVnu}.
There have been more candidates  suggested for dark matter after the X-ray line was identified \cite{models}.  

In this work, we consider a decaying dark matter model with two Majorana fermions. Similarly to the sterile neutrino case, the X-ray line can be obtained from the decay of the heavier Majorana fermion, for a sufficiently small magnetic transition dipole moment and a mass difference of $3.55\,{\rm keV}$ between the two Majorana fermion masses. To this purpose, we propose a microscopic model for dark matter containing the fermion-portal couplings with a new charged scalar \cite{ibarra,bai}. 
We introduce an approximate $Z_2$ symmetry for the long-lived dark matter and a $U(1)_X$ global symmetry for controlling the $Z_2$ breaking to a small amount. As a consequence, a fermion-portal coupling for the heavier Majorana fermion preserves $Z_2$ and is sizable while a tiny coupling for the lighter Majorana fermion is produced after an explicit breaking of $Z_2$. 
Then, when fermion-portal couplings break CP,  a tiny transition magnetic moment between two Majorana fermions is degenerated at one loop \cite{RHneutrino}, even for the charged scalar of weak scale mass.
Therefore, the heavier Majorana fermion is sufficiently long-lived for explaining the X-ray line and  it  can be thermally produced due to the sizable fermion-portal coupling to the SM charged fermion.

We study the parameter space for the masses and couplings of dark matter and the charged scalar, that are consistent with both the X-ray line and the dark matter relic density in our model.  
We also discuss the phenomenological constraints on the model, coming from indirect and direct detection experiments, precision measurements such as the anomalous magnetic moment of muon, and collider experiments.

The paper is organized as follows.
We begin with the properties of magnetic dark matter explaining the X-ray line and describe a microscopic model for the magnetic dark matter in a simple extension of the SM with a charged scalar.
Then, we impose the condition for the relic density in our model and discuss the compatibility with the X-ray line signal. Next, various phenomenological constraints on the model are given.
Finally, conclusions are drawn.

\section{Magnetic dark matter and X-ray line}

We consider a magnetic dipole operator for Majorana fermion dark matter and discuss the required properties for explaining the cluster X-ray line at $3.55\,{\rm keV}$.

For two singlet Majorana fermions, $\chi_1$ and $\chi_2$, having masses $m_{\chi_1}$ and $m_{\chi_2}$,  respectively, we introduce an effective  transition magnetic operator between them as
\be
{\cal L}= \frac{ m_{\chi_2}}{\Lambda^2} \, {\bar\chi}_2 i \sigma^{\mu\nu}  \chi_1 F_{\mu\nu}  +{\rm h.c.} \label{mdm2}
\ee
with $\Lambda$ being the effective cutoff scale.
Then, for $m_{\chi_2}>m_{\chi_1}$, the decay rate of the heavier Majorana fermion into the light one and a photon is
\bea
\Gamma(\chi_2\rightarrow \chi_1+\gamma)&=&\frac{m^5_{\chi_2}}{2\pi \Lambda^4} \left(1-\frac{m^2_{\chi_1}}{m^2_{\chi_2}}\right)^3 \nonumber \\
&=&\frac{4 m^2_{\chi_2}}{\pi \Lambda^4} \,E_\gamma^3
\eea
where in the second line, use is made of the photon energy given by 
\be
E_\gamma=\frac{1}{2}m_{\chi_2}\left(1-\frac{m^2_{\chi_1}}{m^2_{\chi_2}}\right). 
\ee
Thus,  the larger the dark matter mass, the more tuning we need between Majorana masses for $E_\gamma=3.55\,{\rm keV}$. For instance, for $m_{\chi_2}=10\,{\rm GeV}$, we need $|\Delta m /m_{\chi_2}|= 3.55\times10^{-7}$ with $\Delta m\equiv m_{\chi_2}-m_{\chi_1}$.
For a $3.55\,{\rm keV}$ dark matter, the necessary value of the lifetime of dark matter for the X-ray line is $\tau_{\rm DM}=0.20$--$1.8\times 10^{28}\,{\rm sec}$ \cite{xray}, which is equivalent to $\Gamma_{\rm DM}=0.36$--$3.3 \times 10^{-52}\,{\rm GeV}$.
For $m_{\chi_2}=10\,{\rm GeV}$, assuming that $\chi_2$ occupies the whole dark matter relic density \footnote{If $\chi_1$ contributes to the dark matter relic density too, the lifetime of the $\chi_2$ particle must be smaller so that the suppression scale becomes smaller. },  the necessary lifetime of dark matter is rescaled to $\tau_{\chi_2}=0.14$--$1.3\times 10^{22}\,{\rm sec}$ and the decay width of dark matter is $\Gamma_{\chi_2}=0.51$--$4.6\times 10^{-46}\,{\rm GeV}$, so
the required suppression scale of the magnetic dipole operator is then given by
$\Lambda= (0.59-1.0)\times10^{8}\,{\rm GeV}$.

When $\chi_1=\nu$ is the SM neutrino and $\chi_2=\nu_s$ is a sterile neutrino,  a large suppression factor necessary for the X-ray line can be attributed to a small Yukawa coupling for the sterile neutrino. It has been shown that the $3.55\,{\rm keV}$ X-ray line can be obtained for a small mixing angle between the SM neutrino and the sterile neutrino,  $\theta^2\simeq \frac{m_{\nu}}{m_s}\sim 10^{-11}$,
and the sterile neutrino mass, $m_s=7.1\,{\rm keV}$ \cite{xray}.

\section{Microscopic origin of magnetic dark matter}

In this section, we consider a simple model for generating a tiny magnetic dipole moment for dark matter discussed in the previous section.

The minimal setup to obtain the magnetic dipole operator for dark matter is to introduce only a charged scalar $\phi$, that couples between two Majorana singlet fermions, $\chi_1$ and $\chi_2$, and the SM $SU(2)_L$-singlet charged fermion $\psi_R$, the so called fermion-portal couplings \cite{ibarra,bai}, as follows,
\bea
-{\cal L}_{\rm DM}&=& (\epsilon {\bar\psi}P_L{\chi}_{1}  \phi +\lambda {\bar\psi}P_L{\chi}_{2} \phi+ m_\psi {\bar\psi}_R \psi_L+{\rm c.c.})  +m^2_\phi |\phi|^2 \nonumber \\
&&+\frac{1}{2}m_{\chi_1}\overline {\chi_{1}} \chi_{1}+\frac{1}{2}m_{\chi_2}\overline {\chi_{2}} \chi_{2}+\left(\frac{1}{2}\delta \,\overline {\chi_{1}} \chi_{2} +{\rm c.c.} \right)\label{model2}
\eea
where the electromagnetic charges are $q_{\psi_R}=q_{\phi}=-1$ for charged leptons,  $q_{\psi_R}=q_{\phi}=+\frac{2}{3}$ or $-\frac{1}{3}$ for up or down-type quark, and $\psi_L$ is the left-handed part of the SM charged fermion belonging to an $SU(2)_L$ doublet.

We introduce a $Z_2$ discrete symmetry under which $\chi_2$ and $\phi$ are odd while $\chi_1$ is even. 
In this case, we get $\epsilon=\delta=0$ in the Lagrangian (\ref{model2}). 
Moreover, we add a $U(1)_X$ global symmetry under which $\chi_1\rightarrow e^{i\alpha\gamma^5}\chi_1$, $\chi_2 \rightarrow e^{-i\alpha\gamma^5}\chi_2$, and $\phi\rightarrow e^{i\alpha}\phi$  while $\psi$ does not transform. Then, the tree-level Majorana masses for $\chi_1$ and $\chi_2$ are forbidden, that is, $m_{\chi_1}=m_{\chi_2}=0$.
We also introduce a singlet complex scalar $S$ that transforms as $S\rightarrow e^{-2i\alpha} S$ under the $U(1)_X$ global symmetry and is $Z_2$-even. The hyper charges, $U(1)_X$ charges and $Z_2$ parities are summarized in Table 1. 

\begin{table}[ht]
\centering
\begin{tabular}{|c||c|c|c|c|c|}
\hline 
& $\psi_R$ & $\chi_{1L}$ & $\chi_{2L}$ &  $\phi$ & $S$   \\ [0.5ex]
\hline 
$U(1)_Y$ & $q_{\psi_R}$ & $0$ & $0$  & $q_{\psi_R}$ & $0$ \\ [0.5ex]
\hline 
$U(1)_X$ & $0$ & $+1$ & $-1$  & $+1$ & $-2$ \\ [0.5ex]
\hline 
$Z_2$ & $+$ & $+$ & $-$  & $-$ & $+$ \\ [0.5ex]
\hline
\end{tabular}
\caption{Hypercharges, $U(1)_X$ charges and $Z_2$ parities in our model.}
\label{table:charges1}
\end{table}

After the $S$ scalar gets a nonzero VEV and then the $U(1)_X$ symmetry is broken spontaneously, we obtain the Majorana masses as well as the Yukawa coupling as follows,
\be
-{\cal L}_{\rm S}= \frac{1}{2}y_1S {\bar\chi}_1 P_L \chi_1 +\frac{1}{2}y_2 S^* {\bar\chi}_2 P_L \chi_2 +\frac{\kappa}{\Lambda_{UV}}\,S{\bar\psi}P_L{\chi}_{1}  \phi  +{\rm c.c.}  \label{saction}
\ee
where the last term preserves the $U(1)_X$ but it breaks the $Z_2$ symmetry explicitly at the UV cutoff scale $\Lambda_{UV}$ so it is the source for making dark matter to decay. Then, for $\langle S\rangle\neq 0$, we get $m_{\chi_{1,2}}=y_{1,2}\langle S\rangle$ and $\epsilon=\kappa \langle S\rangle/ \Lambda_{UV}$. For instance, for $\langle S\rangle=100\,{\rm GeV}$ $(100\,{\rm TeV})$, $\kappa=0.1-1$ and $\Lambda_{UV}=10^{8}$--$10^{11}\,{\rm GeV}$ ($10^{11}$--$10^{14}$\,{\rm GeV}), we get $|\epsilon|\sim 10^{-9}$--$10^{-7}$ and $m_{\chi_{1,2}}\sim 100\,{\rm GeV}$ for $y_1\sim y_2\sim 1 (10^{-4})$. Henceforth, having in mind the above microscopic model with an approximate $Z_2$ symmetry, we assume that $\delta=0$ and $|\epsilon|\ll 1$ in the Lagrangian (\ref{model2}).

We note that there exists a Goldstone boson (the imaginary part of $S$) after the $U(1)_X$ symmetry is broken spontaneously.
The Goldstone boson can get mass due to QCD-like anomalies in the hidden sector or an explicit breaking of the $U(1)_X$ symmetry. For instance, for $\langle S\rangle \sim 100\,{\rm TeV}$ and the hidden QCD scale $\Lambda\sim 10\,{\rm TeV}$, one can obtain the mass of the Goldstone boson by $m_a\sim \frac{\Lambda^2}{\langle S\rangle}\sim 1\,{\rm TeV}$. The Goldstone boson couples to the SM fermion via a higher dimensional operator in eq.~(\ref{saction}). As the charged scalar decays mostly to a SM fermion and dark matter by fermion portal coupling, the Goldstone interaction to the SM could not be seen at the collider. But, we will comment on the effect of the Goldstone interaction on the cosmological predictions for dark matter in the next section.

\begin{figure}[tb]
\centering
\includegraphics[width=7cm]{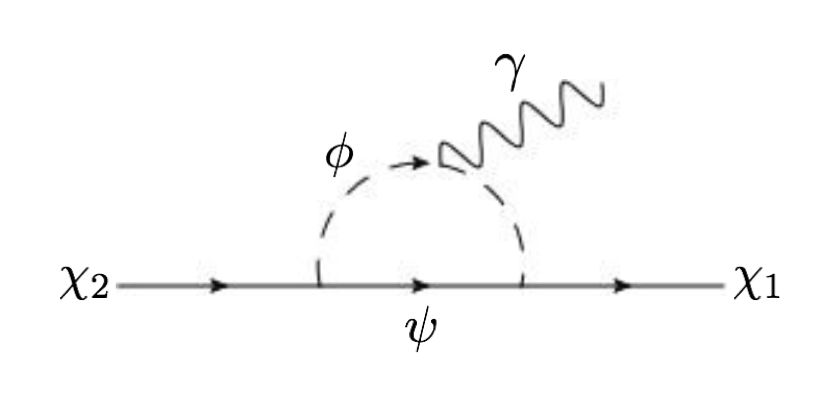} 
\includegraphics[width=7cm]{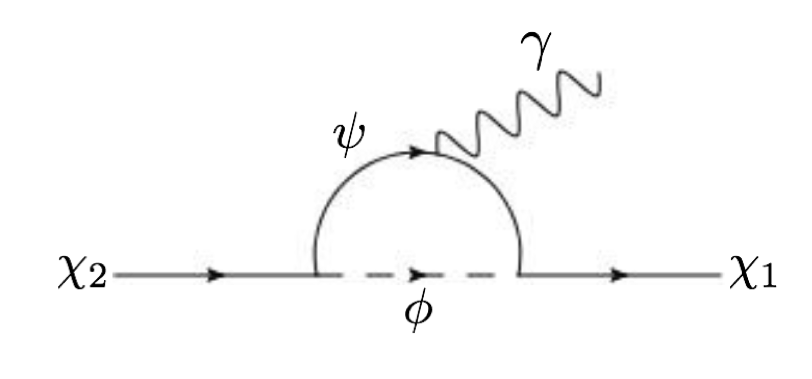}
\caption{Feynman diagrams relevant for the decay of dark matter.
}
\label{fig:decay}
\end{figure}

Computing the one-loop corrections and taking $m_{\chi_2}\approx m_{\chi_1}$, we obtain the effective transition magnetic moment operator for dark matter as
\bea
{\cal L}_{mdm} &=& \frac{ e f_\chi }{2m_{\chi_2}}\,{\bar{\chi}}_2 i\sigma^{\mu\nu} {\chi}_1 F_{\mu\nu}
\eea
where
\bea
f_\chi &=&N_c\, \frac{{\rm Im}(\epsilon^*\lambda)}{16\pi^2} \, m^2_{\chi_2} \bigg[ \int^1_0 dx\,\frac{q_\phi x^2(1-x)}{m^2_{\chi_2}x^2+(m^2_\phi - m^2_{\chi_2})x+m^2_\psi (1-x) }  \nonumber \\
&&+ \int^1_0 dx\,\frac{q_{\psi_R} x^2(1-x)}{m^2_{\chi_2}x^2+(m^2_\psi - m^2_{\chi_2})x+m^2_\phi (1-x) }  \bigg]  \nonumber \\
&=&N_c\, q_\phi\,\frac{{\rm Im}(\epsilon^*\lambda)}{16\pi^2} \, m^2_{\chi_2} \int^1_0 dx\,\frac{ x(1-x)}{m^2_{\chi_2}x^2+(m^2_\phi - m^2_{\chi_2})x+m^2_\psi (1-x) }  \nonumber \\
&\approx& N_c\, q_\phi\, \frac{{\rm Im}(\epsilon^*\lambda)}{32\pi^2} \, \frac{m^2_{\chi_2} } {m^2_\phi}
\eea
where in the last line we assumed that $m_\phi\gg m_{\chi_{1,2}},m_\psi$ and $N_c$ is the number of colors, which is also included for the case of a colored scalar. 
Even if the Dirac mass vanishes, that is, $\delta=0$ in eq.~\ref{model2}, due to the $Z_2$ symmetry in our model, the transition magnetic moment does not vanish because of the $CP$ violating fermion-portal couplings \cite{RHneutrino}, meaning that ${\rm Im}(\epsilon^*\lambda)\neq 0$.
We also note that  ${\bar{\chi}}_1 \sigma^{\mu\nu} {\chi}_1$ and ${\bar{\chi}}_2 \sigma^{\mu\nu} {\chi}_2$  are identically zero for Majorana fermions.

Consequently, as compared to the effective operator introduced in eq.(\ref{mdm2}),  we can  identify the suppression scale of the transition magnetic moment operator as
\bea
\Lambda=\sqrt{\frac{64\pi^2}{e |q_\phi | N_c} }\,\frac{m_\phi}{\sqrt{|{\rm Im}(\epsilon^*\lambda)|}}. \label{cutoff}
\eea
If $\epsilon$ and/or $\lambda$, are small, we can allow for the charged scalar  to be much lighter than the naive cutoff estimated from the effective operator in the previous section. For instance, for $\sqrt{|{\rm Im}(\epsilon^*\lambda)|}\sim 10^{-4}$, $m_\phi$ can be of order $100\,{\rm GeV}$, which will be required for obtaining the relic density from the thermal freezeout for $\lambda={\cal O}(1)$. 
Therefore, for $\lambda={\cal O}(1)$, we need $|\epsilon|\sim 10^{-8}$ for the phase difference of order 1 between the Yukawa couplings \footnote{We note that for a small phase difference, the $\epsilon$ coupling can be sizable too. But, in our model, a small $\epsilon$ is favored as it breaks the $Z_2$ symmetry.}.  As discussed previously, such a small $\epsilon$ and weak-scale Majorana masses for $\chi_1$ and $\chi_2$ can be obtained from the breaking of a $U(1)_X$ global symmetry. 

We remark on the mass difference between the Majorana fermions in our model. 
First, a $Z_2$ breaking Dirac mass term, ${\bar \chi}_1\chi_2$, which could split Majorana masses, is generated under the loop corrections but it is sufficiently small due to a tiny $\epsilon$. 
Second, the fermion portal couplings, $\lambda$ and $\epsilon$, in eq.~(\ref{model2}), do not generate one-loop Majorana masses.
On the other hand, the Yukawa couplings of the singlet scalar $S$, $y_1$ and $y_2$, in eq.~(\ref{saction}), generate Majorana masses at one-loop. For $m_{a,s}\gg m_{\chi_{1,2}}$ where $m_{a,s}$ are masses of singlet pseudo-scalar and scalar of $S$, the one-loop Majorana masses are approximated as
\bea
\delta m_{\chi_1}&\approx&\frac{y^2_1 m_{\chi_1}}{128\pi^2} \,\ln\left(\frac{m^2_s m^2_a}{m^4_{\chi_1}}\right), \\
\delta m_{\chi_2}&\approx&\frac{y^2_2 m_{\chi_2}}{128\pi^2} \,\ln\left(\frac{m^2_s m^2_a}{m^4_{\chi_2}}\right).
\eea
Therefore, the mass splitting at one loop is given by
\be
\delta m_{\chi_2}- \delta m_{\chi_1}\approx \frac{1}{128\pi^2} \,(y^2_2 m_{\chi_2}-y^2_1 m_{\chi_1})\, 
\ln\left(\frac{m^2_s m^2_a}{m^4_{\chi_1}}\right)
\ee 
Consequently, for $y_1\sim y_2$ and $m_{\chi_2}-m_{\chi_1}\approx E_\gamma=3.55\,{\rm GeV}$, 
the splitting between Majorana masses remains small for explaining the X-ray line,
although there is a need of fine-tuning to keep almost degenerate Majorana masses at tree level.

\section{Relic abundance}

\begin{figure}[tb]
\centering
\includegraphics[width=10cm]{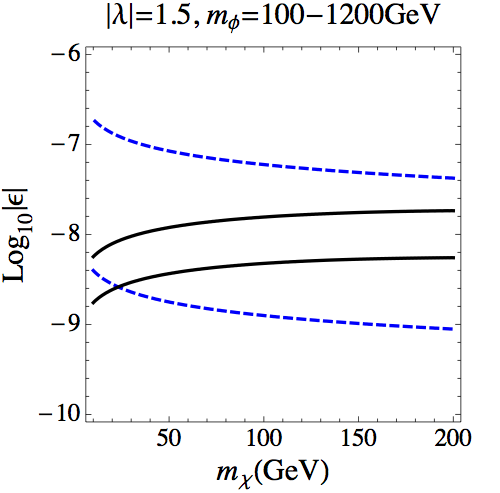} 
\caption{Constraints on dark matter mass vs $Z_2$-breaking coupling from the relic density (Planck $5\sigma$ band) and the X-ray line. The region allowed by X-ray line only is between the blue dashed lines while the region satisfying both X-ray line and relic density is between black solid lines.  We have taken  $q_\phi=-1$. The charged scalar mass is taken between $100\,{\rm GeV}$ and $1200\,{\rm GeV}$ for the X-ray line, but it is restricted to $m_\phi=115-455\,{\rm GeV}$ by the relic density within Planck $3\sigma$ band.}
\label{fig:bounds}
\end{figure}

In this section, we discuss the thermal production of a decaying dark matter in our model and the compatibility with the X-ray line.

From eq.~(\ref{model2}),  the heavier Majorana state couples sizably to the SM charged fermion, unlike the lighter Majorana state, whose coupling is suppressed by $\epsilon$. Then, the t-channel diagram containing the charged scalar is dominant and it gives rise to a sizable annihilation cross section of ${\chi}_2$, that is appropriate for determining the right thermal relic density for dark matter.  The coannihilation between two Majorana states is suppressed by a tiny coupling to the lighter Majorana fermion.
On the other hand, the lighter Majorana state, ${\chi}_1$, couples very weakly to the SM fermion so it cannot be a thermal dark matter.  
We note that when the neutral scalar $S$ introduced in eq.~(\ref{saction}) is light and mixes with the SM Higgs, it could lead to additional contributions to the annihilation processes of $\chi_1$ and $\chi_2$.  
But, we don't consider this possibility for simplicity. It would be worthwhile to investigate the effect of the Higgs-portal interaction of the neutral scalar $S$ in a separate work \cite{progress}.

For the velocity times cross section of dark matter annihilation, $\sigma v=a+b v^2$,  the dark matter relic density is given by
\begin{equation}
\Omega_{\rm DM}h^2=\frac{2.09\times 10^8\,{\rm GeV}^{-1}}{M_{Pl}\sqrt{g_{*s}(x_F)}\,\left(a/x_F+3b/x^2_F\right)} \, ,
\end{equation}
where the freeze-out temperature gives $x_F=m_{\rm DM}/T_F\approx20$ and $g_{*s}(x_F)$ is the number of the effective relativistic degrees of freedom entering in the entropy density.

In our model, the velocity times annihilation cross section for ${\chi}_2 {\chi}_2\rightarrow \psi {\bar \psi}$ is p-wave and temperature-suppressed \cite{ibarra,bai} as
\be
\langle \sigma v\rangle= \frac{N_c {|\lambda|}^4}{16\pi}\,\frac{m^2_{{\chi}_2} (m^4_{{\chi}_2}+m^4_\phi)}{(m^2_{{\chi}_2}+m^2_\phi)^4} \cdot\,\frac{T}{m_{\chi_2}}.
\ee
Nonetheless, the annihilation cross section can be sufficiently large for thermal dark matter.
In Fig.~2 and 3, the parameter space for dark matter couplings and mass parameters satisfying the relic density and the dark matter decay rate required for the X-ray line is shown in black and in blue, respectively. We imposed the $3\sigma$ band of $\Omega_\chi h^2=0.1199\pm 0.0027$ from Planck \cite{planck} and used  $\Gamma_{\chi_2}=0.36$--$3.3 \times 10^{-52}\,{\rm GeV} \,(m_{\chi_2}/3.55\,{\rm keV})$ for the decay rate of dark matter.  

In Fig.~2, the $Z_2$ breaking fermon-portal coupling $|\epsilon|$ compatible with the X-ray line lies between $10^{-9}$--$10^{-7}$ for weak-scale masses for dark matter and charged scalar. The larger the $\epsilon$ coupling, the larger the charged scalar mass for obtaining the X-ray line and the larger the dark matter mass for explaining the relic density as well.
Thus, the relic density condition constrains charged scalar and dark matter masses as well as the $\epsilon$ coupling further.  In the example with $|\lambda|=1.5$, charged scalar mass ranges between $115\,{\rm GeV}$ and $455\,{\rm GeV}$ for $m_{\chi_2}=10-200\,{\rm GeV}$, and there can be a lower bound on dark matter mass from the X-ray line.  In the case of a colored scalar with $|\lambda|=1.5$, the X-ray line and the relic density can be obtained at the same time, but the colored scalar mass is bounded by the relic density to the values between $152\,{\rm GeV}$ and $647\,{\rm GeV}$  for $m_{\chi_2}=10-200\,{\rm GeV}$. When dark matter mass is larger than $200\,{\rm GeV}$, the colored scalar can be as heavy as  about $800\,{\rm GeV}$ so it is compatible with the current LHC limits as will be discussed in Section 5.2.

In Fig.~3, for the fixed $|\epsilon|=10^{-8}$, we varied the $Z_2$ preserving fermion-portal coupling $|\lambda|$ and weak-scale mass parameters, resulting in a wide range of the consistent values of the parameters for $|\lambda|={\cal O}(1)$. 
On the left panel of Fig.~3, both the X-ray line and the relic density can be satisfied for $|\lambda|=1.3-2.0$ for $m_\phi=350\,{\rm GeV}$ and $|\epsilon|=10^{-8}$.
On the right panel of Fig.~3, the lower bound for dark matter mass is given by about $30\,{\rm GeV}$, which is consistent with Fig.~2 for $|\epsilon|=10^{-8}$.

\begin{figure}[tb]
\centering
\includegraphics[width=7.5cm]{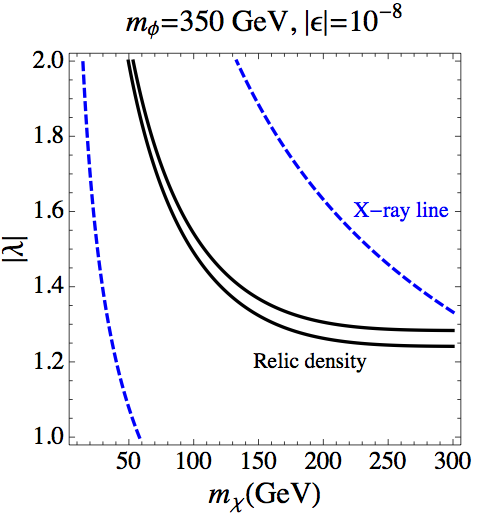} 
\includegraphics[width=7.5cm]{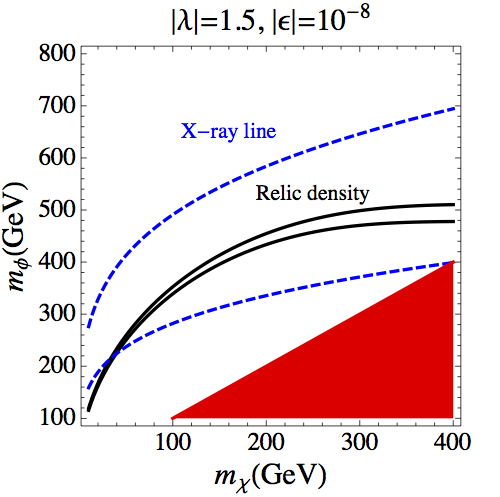}
\caption{Constraints on the dark matter coupling and charged scalar and dark matter masses from the relic density (Planck $3\sigma$ band) and the X-ray line. The region allowed by Planck (X-ray line) is between the black solid lines (blue dotted lines).
We have taken  $q_\phi=-1$, $|\epsilon|=10^{-8}$. The red area is excluded because the charged scalar is lighter than dark matter.
}
\label{fig:bounds}
\end{figure}

Before closing the section, we remark on the effects of the singlet scalar $S$ in our model discussed in the previous section.
First, if the VEV of the singlet scalar $S$ is of order weak scale, the singlet real scalar of $S$ can be light and its Yukawa couplings of Majorana fermions, $y_1$ and $y_2$, can be sizable. In this case, while the Majorana fermion $\chi_2$ was in thermal bath with the SM fermions due to the fermion portal coupling $\lambda$, its Yukawa coupling to $S$ could thermalize the Majorana fermion $\chi_1$ by the scattering process of $\chi_2$ to $\chi_1$. In particular, the Goldstone boson of $S$ could contribute dominantly to this process. As a consequence, a resultant non-negligible abundance of $\chi_1$ could require the decay rate of $\chi_2$ for the X-ray line to be larger than in the case where $\chi_2$ explains the full relic density of dark matter. 
On the other hand, in the case of a relatively large singlet VEV, for instance, $\langle S\rangle=100\,{\rm TeV}$, the Yukawa couplings, $y_1$ and $y_2$, are of order $10^{-4}$, for weak-scale Majorana masses.
In this case, for the Goldstone boson mass of $m_a\sim 1\,{\rm TeV}$, the abundance of the Majorana $\chi_1$ produced from the scattering of $\chi_2$ can be small enough \cite{fimp}, so that the current relic density is dominated by $\chi_2$. 
In this work, we assumed the latter case and focused on the effect of the fermion portal coupling $\lambda$ on dark matter. 

We also note that the Higgs portal coupling to the singlet scalar, such as $\lambda_{sh} |S|^2 |H|^2$ where $H$ is the Higgs doublet, may make dark matter to annihilate into the SM particles through the singlet scalar and affect the cosmological predictions for dark matter in our model too. Importantly, the Majorana fermion $\chi_1$ can annihilate into a pair of the SM particles throughout the Higgs portal coupling too, so it could be another component for thermal dark matter.  In this case, the relic density of the decaying dark matter $\chi_2$ would be smaller so the decay rate of $\chi_2$ should be larger to accommodate the X-ray line. 
If the mixing angle between the Higgs and singlet scalar is small enough, namely, $\theta\sim \frac{\lambda_{hs} v}{\langle S\rangle}\ll 1$ with $v$ the Higgs VEV,  the Higgs portal coupling would affect our previous discussion for dark matter little, enabling us to focus on the fermion portal coupling.   The detailed discussion on the case with non-negligible mixing angle will be done in a separate work \cite{progress}.

\section{Various constraints}

In this section, we discuss various constraints on the model from indirect and direct detection experiments, precision measurement and collider experiments.

\subsection{Indirect and direct detections}

The annihilation cross section of our decaying dark matter is p-wave suppressed.
Thus, there is no bound on the model from the indirect detection experiments.

If our dark matter couples to the light quarks, dark matter can scatter elastically with nucleons by either tree-level diagram with the charged scalar in the s-channel or t-channel through transition magnetic moment or a box diagram containing the charged scalar. First, the tree-level diagram for Majorana dark matter leads to only a spin-dependent cross section \cite{bai}. So, a sizable spin-independent cross section comes only from the box diagram involving the charged scalar, when dark matter couples to up or down quark \cite{batell,nelson}.
Namely, the up or down quark coupling to dark matter is bounded by direct detection experiments such as XENON100 \cite{xenon100} and LUX \cite{lux}. 
On the other hand, the transition magnetic moment for dark matter is suppressed for a decaying dark matter, resulting in no sizable contribution to direct detection.
The detailed discussion on direct detection will be presented elsewhere \cite{progress}.

\subsection{Bounds on the charged scalar}

When the charged scalar carries electromagnetic charge $q_\phi=-1$, dark matter couples to leptons only and can contribute to the anomalous magnetic moment of the muon at one-loop.  As the heavier dark matter fermion can couple sizably to the muon, the effective magnetic moment operator for the muon is given by
\be
{\cal L}_{\mu}= \frac{e f_\mu }{2m_\mu} \,{\bar\mu}\sigma^{\mu\nu} \mu F_{\mu\nu}
\ee
with
\be
f_\mu =\frac{{|\lambda|}^2}{32\pi^2} m^2_\mu \int^1_0 dx\,\frac{q_\phi x^2(1-x)}{m^2_{\mu} x^2+(m^2_\phi - m^2_{\mu})x+m^2_{\chi_2} (1-x) }. 
\ee
Thus, for $m_\phi\gg m_{\chi_2}, m_\mu$,  there is a negative contribution to the anomalous magnetic moment of the muon as
\be
\Delta a_\mu=-\frac{{|\lambda|}^2}{96\pi^2} \,\frac{m^2_\mu}{m^2_\phi}.
\ee
Therefore, there is a bound on the dark matter coupling from the measurement \cite{mug-2} of the muon $(g-2)_\mu$ as $|\Delta a_\mu|<3.45\times 10^{-12}$, which leads to $|\lambda|<0.27 \,(m_\phi/500\,{\rm GeV})$.
This bound is in a tension with the parameter space satisfying the relic density, when the muon coupling  determines the relic density dominantly. 

The couplings of dark matter to other leptons can be constrained by flavor violating processes. For instance,  the bound on the muon decay process, ${\rm Br}(\mu\rightarrow e\gamma)<5.7\times 10^{-13}$ at $90\%$ C.L. \cite{meg}, constrains the additional coupling in $\lambda_e {\bar e}P_L\chi_2\phi$ to $|\lambda_e|<0.01 (m_\phi/100\,{\rm GeV})$.
On the other hand, the tau coupling is less constrained as ${\rm Br}(\tau\rightarrow \mu(e)\gamma)\lesssim 10^{-8}$ \cite{taudecay} so it could give a dominant contribution to the annihilation and decay of dark matter. 

Now we mention the collider bounds briefly. 
When the charged scalar couples to light quarks, 
it can be produced copiously at the LHC by gluon fusion or quark-gluon interaction, leading to monojet or two jets plus missing energy in the final state \cite{bai}.
When the charged scalar couples to leptons, it can be produced by Drell-Yann process at the LHC or the ILC. The signature is one or two leptons plus missing energy. The upgraded LHC can cover the parameter space relevant for the decaying dark matter saturating the relic density as well as explaining the X-ray line.
According to the current ATLAS and CMS data \cite{lhcbound} , we have excluded the masses of  the charged scalar $90\,{\rm GeV}\lesssim m_\phi\lesssim 325\,{\rm GeV}$ and $110\,{\rm GeV}\lesssim m_\phi\lesssim 280\,{\rm GeV}$, respectively, for leptonic scalars, and 
$m_\phi \lesssim 640(680) \,{\rm GeV}$ and $m_\phi \lesssim 620(700)\,{\rm GeV}$, respectively, for colored scalars such as stop (sbottom) \cite{stopsbottom} but the limits depend on the decaying processes of the charged scalar and the mass gap. On the other hand, LEP leads to a less model-dependent bound, $m_\phi\gtrsim 100\,{\rm GeV}$ \cite{lepbound}.  Depending on the models, the previous collider limits and the LHC Run II can constrain the charged scalar masses explaining the X-ray line. For instance, for $|\lambda|=1.5$, the lepton model with $325\,{\rm GeV} \lesssim m_\phi \lesssim 500\,{\rm GeV}$ or the quark model with $640\,{\rm GeV}\lesssim m_\phi \lesssim 800\,{\rm GeV}$ can be searched for in the LHC Run II.

\section{Conclusions}

We have studied a simple model of dark matter with two singlet Majorana fermions, explaining the X-ray line as the decay product of dark matter through a transition magnetic moment.
We showed that a tiny value of the transition magnetic moment is generated at one loop by the interplay between the $CP$ violating fermion-portal couplings of a charged scalar to the two Majorana fermions: one to the heavier state preserves $Z_2$ and the other to the lighter state breaks $Z_2$ by a tiny amount. Therefore, even with a weak-scale charged scalar, the heavier Majorana fermion decays into the lighter one with a sufficiently long lifetime, emitting the X-ray line at $3.55\,{\rm keV}$. Moreover, the decaying Majorana fermion can be a thermal dark matter due the sizable $Z_2$-symmetric coupling to the SM charged fermions.
It was shown that the proposed model can satisfy the current bounds from the relic density and various experiments and it can be testable in the upgraded LHC.

\section*{Acknowledgments}

The author would like to thank Young Min Kim for helpful discussion on the work. 
The work of HML is supported in part by Basic Science Research Program through the National Research Foundation of Korea (NRF) funded by the Ministry of Education, Science and Technology (2013R1A1A2007919).

\end{document}